\documentclass[runningheads]{llncs}
\usepackage{graphicx}
\usepackage{hyperref}
\usepackage{subcaption}
\usepackage{amsmath}
\usepackage[utf8]{inputenc}
 \usepackage{bibnames}
\DeclareMathOperator*{\argmax}{arg\,max}
\DeclareMathOperator*{\argmin}{arg\,min}

\begin{document}
\title{Scalable Algorithmic Infrastructure for Computation of Social Crowding and Viral Disease Encounters - \textit{mContain} Case Study}
\titlerunning{mContain}
%
\author{Md Azim Ullah}
\institute{The University of Memphis}


%
\maketitle
\begin{abstract}
\textit{mContain} was developed (and sparsely deployed) by MD2K center at University of Memphis in the early stages of COVID-19 pandemic to help reduce community transmission in Shelby County and Memphis metropolitan area. The application counts and displays the number of daily proximity encounters with other app users. To reduce the chances of entering crowded places, users can see the level of crowding at busy places on a map. If a user and their COVID-19 test provider both agree to share the results of their test, the app can notify other users about possible exposures to COVID-19. The smartphone application collects location and Bluetooth data and sends it to cloud for near real time processing and decisions to be sent back for visualization and interface with the user. The backend algorithmic infrastructure responsible for real time crowd estimation and contact tracing from streaming batch data use open-source cloud analytics platform Cerebral-Cortex. This project concerns about presenting the authors contributions in the algorithmic development, design and implementation of \textit{mContain} application as part of the entire collaborative project.  We describe the \textit{mcontain} algorithmic infrastructure and major computational challenges encountered when developing and deploying this application for real-life usage. Details of the app can be found in \href{https://mcontain.md2k.org/}{https://mcontain.md2k.org/}.

\end{abstract}

%

%
%
%

\section*{Acknowledgement}

The \textit{mContain} application, software, and website was developed by the software engineers, staff, and doctoral students at the MD2K Center of Excellence at The University of Memphis. Members of the team include Dr. Santosh Kumar, Dr. Manoj Jain, Dr. Timothy Hnat, Dr. Anandatirtha Nandugudi, Dr. Monowar Hossain, Dr. Nasir Ali, Brian Ahern, Joe Biggers, Shahin Samiei, Md Azim Ullah, Rabin Banjade, Soujanya Chatterjee, Shiplu Hawlader, Hosneara Ahmed, Sayma Akther, Sameer Neupane, Mithun Saha \& Dr. Nazir Saleheen. Although the entire project is described here for completeness purposes, author's individual role in the project lies in design, development and implementation of the cloud algorithms for real time computation of proximity encounters and social crowding using Cerebral-Cortex. Cerebral-Cortex is a open source big data processing framework developed by  MD2K. Author would like to acknowledge all the people who contributed immensely to the successful completion of the \textit{mContain} application whilst facing an emergency public health crisis.

\section{Introduction}
The SARS-COV-2, a new member in the family of coronaviruses, was declared a pandemic in 2020. It is a highly infectious virus with a significant mortality risk arising from respiratory failures of infected human carriers. The onset of the pandemic presented major challenges across the scientific community for innovation and use of technology for coming up with effective measures of fighting against a novel virus. Without the wide availability of testing facilities and uncertain timeline of vaccines, the first approach in the fight against COVID around the world has been on containing the community transmission. Governments at local, regional, and national levels released guidelines for starving the virus of new hosts by reducing people-to-people contacts, commonly referred to as social distancing. As social contact can also occur virtually, we use the term proximity encounter to refer to two or more individuals being within physical proximity (i.e., within 6 feet) of each other, for several minutes (e.g., 10 minutes). Several recommendations are still being made on how people interact with one another while being 6 feet apart. This gave rise to the need for a mobile technology that can measure daily proximity encounters both at the individual and at the population level. We developed such an application, called \textit{mContain} for this purpose and launched it for public usage in the Greater Memphis region. While we learned from other apps deployed around the world during COVID-19 outbreak, \textit{mContain} is distinguished by its focus on being a personal
tool for measuring and tracking personal proximity encounters, issuing personal crowding alerts, aggregating population data to highlight crowding hot-spots, and an anonymous, but, authenticated approach to COVID-19 contact tracing.

We describe the design of \textit{mContain} mobile and cloud infrastructure, developed algorithms, and computational decisions involved in efficient and effective deployment across mass population. We highlight computational and privacy challenges that can improve the accuracy, efficacy, privacy, and overall social utility of apps like \textit{mContain}. These challenges can formulate a research agenda for the computing community looking for ways to contribute to controlling a pandemic and make the world better prepared for future viral disease outbreaks. 

\textbf{Section~\ref{sec:outbreak} through Section~\ref{sec:system_architecture} gives an overview of the complete project to present an understanding of the decision making process involved in development of the application by all contributing members. Section~\ref{sec:algorithms} describes the development of algorithms, design and implementation decisions by the author as part of the project.}

\section{COVID-19 Outbreak - Need for Social Distancing \& Contact Tracing}
\label{sec:outbreak}
The incubation period for COVID-19 ranges
from 1-14 days, which means an individual may
begin to show signs of infection anytime between
1 to 14 days after being exposed to the pathogen. However, even in incubation, the disease is capable of spreading, as evidence suggests 50\% of the transmission occurs within this period through COVID-19 positive patients, who inadvertently become the sources of infection for others coming in close contact. People can also get infected by touching a surface or object that has the virus on it
and then touching their faces without adequately
washing their hands. The severity of illness, though, varies from patient to patient depending on age and other pre-existing conditions. Older adults in particular, and those suffering from medical conditions such as heart disease, diabetes, lung diseases are more likely to develop severe complications from the
exposure. Besides, for a novel virus like this, no vaccine was readily available. Despite the extensive efforts from many countries, as of March 2020 the earliest projection for getting a vaccine out for mass people was not until the second half of 2021. Fortunately, through incredible collective efforts, several vaccines have emerged with ground breaking results. In the absence of vaccine and effective drugs at the onset of pandemic, health care facilities across the world faced an uphill task of accommodating and treating an influx of COVID-19
patients. Against this backdrop, health experts urged people who may not have been exposed
yet, to exercise social distancing by staying at
least six feet away from each other when they
are outside going about their daily activities and avoid crowded places. However, for those who
may have been in close contact with a COVID-
19 patient, the instruction is to practice self-quarantine by staying inside the household for 14 days, strictly maintaining standard hygiene, social distancing and no sharing policies.

Thus to contain the spread of a contagious
disease, it is imperative to identify individuals, mainly those recently exposed, to contact, follow up and monitor their condition for conducting tests as and when required. This process is termed as contact tracing. As robust as it may sound, the downside of traditional manual contact tracing is speed or rather the lack of it in staying ahead of a highly contagious disease, to restrain its
transmission. The other imposing challenge is to
identify the maximum number of contacts which
by no means is a mean feat. Additionally, it is
not enough just to isolate individuals but to raise public awareness about their proximal encounters and crowded places, so that they can play their due role in containing the disease. All of these issues can be addressed more effectively through technology by reaching out to a large number of people instantaneously, making them aware of the situation and possible implications. 

\subsection{Efficacy of Contact Tracing \& Social Distancing}
Contact tracing has been referred to as the linchpin of epidemic control~\cite{sacks2015introduction}. Before COVID-19 outbreak,  past use contact tracing  for effective containment of disease include the tuberculosis outbreak in Botswana~\cite{ha2016evaluation}, Ebola outbreak in Sierra Leone~\cite{danquah2019use}, Guinea~\cite{dixon2015contact}, Liberia~\cite{swanson2018contact}, and also for HIV containment~\cite{armbruster2007contact}. Contact tracing has also been used for other diseases such as severe acute respiratory syndrome
(SARS)~\cite{lipsitch2003transmission}, foot-and-mouth-disease~\cite{kiss2005disease}, smallpox~\cite{porco2004logistics}, and avian influenza~\cite{wu2006reducing}.
The impact of control strategies such as contact tracing or social distancing is characterized by the change in effective reproduction number of a virus ($R_e$) with such measures employed~\cite{kretzschmar2020impact}. With the Ebola epidemic as the primary case study, researchers have shown that increasing contact tracing efforts has the potential to eliminate the disease at a later stage in the outbreak with incremental reduction in $R_e$~\cite{browne2015modeling}. Researchers have shown for COVID - 19 that the effective reproduction number of the virus falls from 1.2 to 0.8 with addition of contact tracing measures~\cite{kretzschmar2020impact} alongside physical distancing. Meeting the exponential demands on emergency health care in times of pandemic also requires limiting the reproduction and transmission capabilities of the virus~\cite{hernandez2020evaluating}. In United Kingdom, social distancing measures were introduced only after researchers projected an increased mortality rate of 460,000 and advocated the use of social distancing~\cite{ferguson2020impact} to reduce the mortality rate and demands on emergency health care. Benefits of social distancing measures have also been quantified in economic terms~\cite{thunstrom2020benefits}. 
Thus, in the face of a novel virus such as COVID-19 control measures such contact tracing and social distancing can play an important role in containing pandemic across the world. 

\subsection{Mobile Health as a Means for Effective Virus Control}
Mobile health technologies have been successfully used for effective Ebola surveillance and Contact Tracing using GPS traces~\cite{sacks2015introduction}. The low resource settings of mobile health applications and their wide availability enable increased access to data sources, effective management of health records and timely health interventions compared to traditional paper based tracing. However, rapid adoption remained a challenge and had not replaced the paper based system wholly due to limitations in hardware availability, fear of stigma and other reasons. Although smartphone-based contact tracing presents a
viable solution to limiting viral transmission, privacy concerns related to the personal data security remain a challenge~\cite{yasaka2020peer}.
Moreover, using location data alone is not sufficient to accurately estimate viral exposure~\cite{farrahi2014epidemic}. Furthermore, the modality of information sharing incorporated in commonly used location sharing apps may not be directly applicable to the healthcare field where patients generally want to remain unidentifiable for fear of social stigma. Thus, development of a working mobile technology based contact tracing application with respect to user privacy is a natural step in the response to the pandemic. 

\subsection{Applications Deployed Worldwide in Response to Novel Coronavirus}

Many countries so far have come up with their
versions of contact tracing apps to counter the
threat of COVID-19. For example, South Korea
developed an app that tracks a COVID-19
patient’s history of visited places and broadcasts alerts about the infected individual’s gender, age, date of infection and place of encounters with others~\cite{hamilton_2020}. Israel launched HAMAGEN~\cite{gov.il} that cross-checks users’ locations traces with that of infected individuals’ to notify users about their possible encounters with them in the past 14 days.
Russia introduced a face recognition app that
uses video footage and location traces to prevent people from going outside during the quarantine period~\cite{ilyushina_2020}. NextTrace~\cite{sharonbegleyapril_2020} is an app that aims to locate COVID-19 patients whenever they are tested positive and notify their contacts instantaneously for helping them to
isolate and self-quarantine without further delay. India used Aarogya Setu~\cite{mitter_2020} to notify a person when he comes within six feet of a COVID-
19 patient. It leverages a government-provided
database to obtain location traces of COVID-19
patients and later on tracks their close contacts using GPS and Bluetooth. To use the app, users have to provide personal information which the app claims to share only with the government of India. TraceTogether~\cite{holmes_2020} is a Bluetooth-based privacy-aware app that tracks individuals who spend at least 30 minutes within 2 meters of an infected individual. It requires COVID-19 patients to share their information with the app so that it can find relative distances among users for tracing close contacts. In the USA, Social Distancing Scoreboard~\cite{rss} assigns grade and
letter to each state and county based on their
performance in maintaining social distancing. It
collects information from John Hopkins Data
Repository to do analysis and display findings of crowding not personalized to any user, interactively. Apple released a screening tool~\cite{apple_newsroom_2021} which allows users to answer a series of questions before recommending either a test, visit a health care provider or self-isolation. It maintains the privacy of users and does not collect any personal data.

\section{mContain - Privacy Constrained Use of User Location and Proximity from Mobile Sensors}
In the given context, we are specifying the
burning research challenges that evolved during
the development and deployment of \textit{mContain}
that we believe will drive better preparation in
facing similar or more severe outbreaks in future.
\textit{mContain} is a free mobile app to help track
social distancing during the COVID-19 outbreak
to reduce community transmission, with highest
regards to users’ right to privacy. The app aims
to provide utility by efficiently alerting close
contacts of COVID-19 positive patients who may
be at immediate risk and informing app users
in general about crowding hot-spots, their recent
proximal encounters with other app users and
visits to crowded places to raise awareness. Like
other location-tracing apps, it also collects location information of the users using GPS. However, a crucial distinction is that not only they are
being stored anonymously in the server against
unique codes representing app users, but the location traces are also kept off-limits to everyone.
Though it is important that individuals share
their personal information in such trying times,
one should not disregard the long-term risks of
sharing such privacy sensitive information. While
many countries have implemented apps in ways
that are not conducive to preserving privacy,
others like USA, Canada or European countries
have to be very careful about introducing any
such app because of the utmost value attached to
individual privacy. So we had to be very careful
in designing the app so that we strictly preserve
the privacy of our potential app users when we
collect as well as share their information. Notably,
the app will notify a user if s/he had encounters
with an infected individual without revealing any
location or privacy-sensitive information about
that individual and also about his/her recent visit
to a crowded place. Moreover, unlike some other
location tracing apps, to ensure the authenticity of
the test results, the testing agencies will provide
the information about the positive test result of
a COVID-19 patient using the associated unique
code. Additionally, a user will only view the
total number of unique encounters s/he had the
previous day in violation of social distancing and
which places were densely crowded. In short \textit{mContain} contributes to raising awareness about social distancing and contact
tracing in the following ways:

\begin{enumerate}
    
    \item Tracks the location of users anonymously to
provide information about their recent proximity encounters with other app users in violation
of social distancing. Traces close contacts of a
positive tested person if person and the testing
agency both give consent for sharing the test
result with the app. The people who were in
close proximity of the patient will get alerts.

    \item Displays the level of crowding at busy places
on a map to reduce the chances of visiting
those places.
    
    \item Informs the user about his/her recent visit to a crowded place

\end{enumerate}

\subsection{Design Goals}
As in the case of a viral pandemic like
COVID-19, containment is the crucial strategy to
adopt to stop community transmission until the
vaccine is developed. Any app or solution should
provide relevant information to users who can use
those information in social distancing and contact
tracing. We outline the following design goals
that should be incorporated into social distancing
and contact tracing solutions.

\begin{enumerate}
    \item \textbf{Limiting proximity encounters}
    \begin{itemize}
        \item \textbf{Information about recent Proximity encounters}: App needs to provide users information about their daily encounters with other app users.
        \item \textbf{Information about recent visit to a crowded place}: Inform a user about a recent visit and sojourn time in a crowded place to raise awareness about social distancing.
        \item \textbf{Information about crowded places}: Inform users about the crowded places at different times of the day to enable them to make informed decisions about visiting those places.
    \end{itemize}
    
    \item \textbf{Automated contact tracing}
    \begin{itemize}
        \item \textbf{Authenticity}: Authenticity in the context of mContain means the error-free identification of the COVID-19 positive individual. As contact tracing starts with identifying the infected individual, it is imperative to have the accurate identification of the infected individual.
        
        \item \textbf{Anonymity}: Anonymity means the protection of the identity of the app users, especially that of the infected individuals. Identification of infected individuals may give rise to social stigma, discrimination, and may well be subject to physical as well as mental harassment. It may also potentially expose their historical locations, their interactions with any businesses, which may result in boycott and retributions.
    \end{itemize}
    
    \item \textbf{Proper Communication with End User}: 
        \begin{itemize}
            \item \textbf{Reducing False Alerts}: Although various other applications generate alerts in different situations such as natural disaster, weather, and traffic condition and so on, types of may differ depending on what they are being used for. It is essential to have proper identification of types of alerts used and also the reliability of the alerts generated. In the context of contact tracing and social distancing, there is not much room for error in the alert generation as a false alert may result in misinformation and panic.
        \end{itemize}

\end{enumerate}

\section{Mobile Application and Website}

\textbf{\textit{mContain} mobile application and website was developed with multi disciplinary collaborative efforts led by the software development team  and staff of MD2K Center of Excellence at The University of Memphis. We present it here to contrast the applicability of the algorithms developed by the author.}

\begin{figure*}[t!]
    \centering
    \begin{subfigure}{0.23\textwidth}
        \centering
        \includegraphics[height=1.2in]{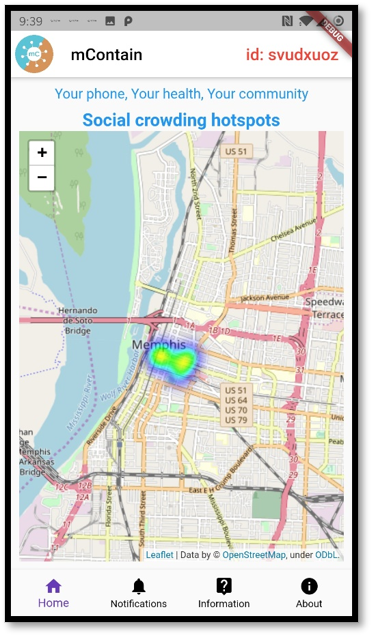}
        \caption{Crowding}
    \end{subfigure}%
    ~ 
    \begin{subfigure}{0.23\textwidth}
        \centering
        \includegraphics[height=1.2in]{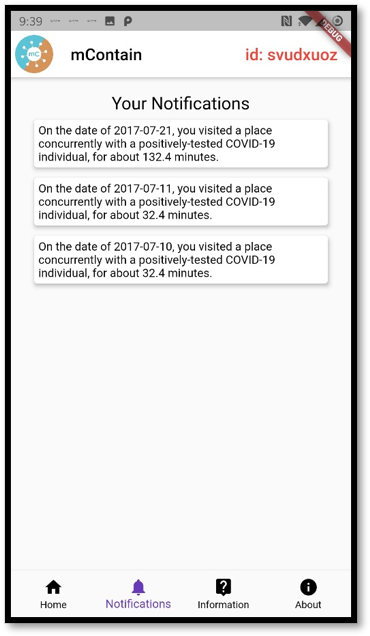}
        \caption{Notifications}
    \end{subfigure}
    ~
    \begin{subfigure}{0.23\textwidth}
        \centering
        \includegraphics[height=1.2in]{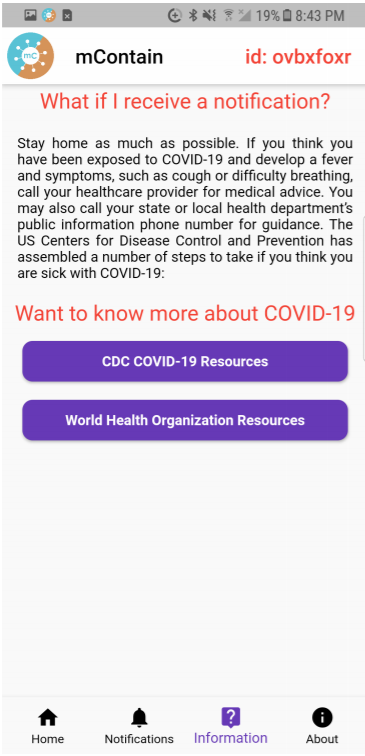}
        \caption{Links}
    \end{subfigure}
    ~
    \begin{subfigure}{0.23\textwidth}
        \centering
        \includegraphics[height=1.2in]{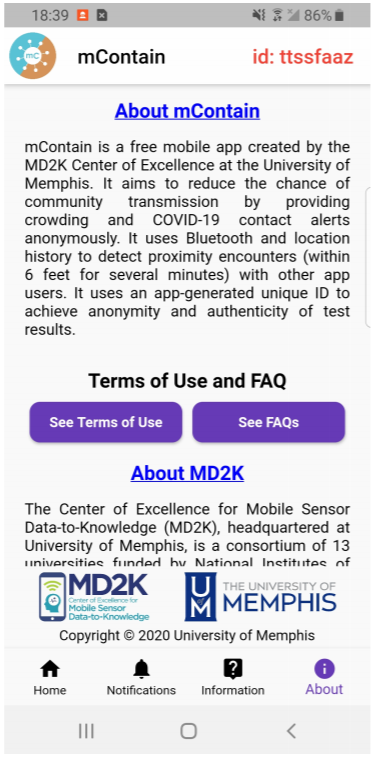}
        \caption{About}
    \end{subfigure}
    \label{fig:mobile_app}
    \caption{mContain Mobile Application}
\end{figure*}

\subsection{Application}
\textit{mContain} app is available for Android devices. While installing the app, we ask users to allow enabling location and Bluetooth services because the phone needs to have these two services running all the time for the best use of the app. There are four different panels in the \textit{mContain} app. The Home panel displays the crowding hot-spots of the previous day throughout the city of Memphis in an interactive map. Definition of crowding hotspot and algorithm for calculating hotspot is discussed below in Section~\ref{sec:algorithms}. Figure~\ref{fig:mobileapp} shows four different panels of the \textit{mcontain} mobile application. A heatmap marks
each crowding hotspot, and the colour of the marker represents the crowd density. Below the map, the app also displays the number of unique encounters with other app users. Notifications panel shows notifications for contact and crowding alerts. Information panel provides the links to the CDC and WHO website for the information and guidelines about COVID-19. About page has the link to \textit{mContain} and MD2K website.

\subsection{Website}
We have also developed \textit{mContain} website where users can go to get more information about the \textit{mContain} app. It provides information about the total number of app users, proximity encounter per user, COVID-19 encounters among users and encounters at the largest crowding event. It also has Frequently Asked Questions section, Terms of Use and privacy notice.The website also provides useful links to CDC and TN Department of Health Information on COVID-19.

\section{System Architecture \& Data Flow}
\label{sec:system_architecture}
\begin{figure}[t]
    \centering
    \includegraphics[width = \textwidth]{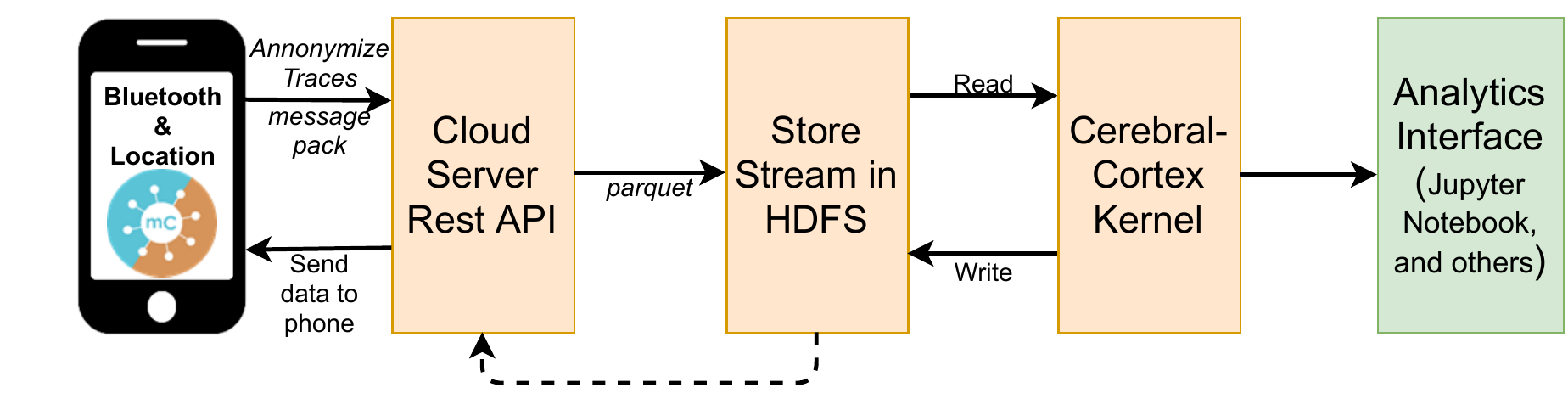}
    \caption{Mobile to Cloud Data Flow in mContain}
    \label{fig:system_architecture}
\end{figure}

Figure~\ref{fig:system_architecture} shows the data flow process for \textit{mContain} from user smartphone to the cloud and presents the top level modular structure of \textit{mContain} cloud. We now describe the bidirectional data flow process with brief descriptions of different modules. \textbf{The cloud architecture and data flow design process was developed by the software development team at MD2K center and continue to be in active use for data collection in multiple mobile health studies across the USA.}

\textit{mContain} collects location and Bluetooth proximity data from users smartphone and uploads it to the cloud using a REST interface. \textit{mcontain} is connected to the REST server through a secure HTTPS protocol. All the data is transferred over HTTPS in a secured and encrypted fashion. Upon installation in mobile phone, the user is provided a 16 digit identifier that displays over the phone application. When the user is registered in the system, another anonymous 32 digit UUID, $u_1$ is generated automatically. This user UUID, $u_1$ is the only directly identifiable information responsible for server side communication to the smart-phone and maps directly to the 16 digit identifier shown in the mobile application. Developers and stakeholders residing on the server side can only access an encrypted version of $u_1$, lets say $u_2$ with $u_2 \neq u_1$. Thus real user identity is protected with multiple layers of security. Once user is registered, the mobile application can upload data to the REST server using pre-specified links allotted for each data stream that the phone wants to upload. Streams are also identified with unique UUIDs. If user wants to upload a data-stream whose UUID is not in the system, the system creates one and provides the user with the UUID information for upload.

The mobile application runs Bluetooth as a foreground process. It broadcasts and listens for Bluetooth beacons from nearby devices running the same software. Beacon exchange between two users is recorded once in every 20 seconds. We also collect one GPS data-point in tandem with every Bluetooth beacon exchange thus limiting the effects of the data collection process on storage and battery usage. The phone software offloads both GPS and Bluetooth data in chunks once every 15 minutes. The offload routine is done in a near real time fashion using \textit{msgpack} binary protocol. \textit{msgpack} is an open source binary serialization format and allows for more efficient data representation compared to alternatives such as JSON and XML and is less prone to corruption. For continuously running Bluetooth in the foreground, we use the Bluetooth Low Energy (BLE) platform~\cite{mackensen2012bluetooth}, supported by Android devices for its lower energy consumption profile. Once REST-API server receives data from smart-phone, it converts the msgpack file into parquet and stores it into HDFS (Hadoop Distributed File System). 

For interfacing to data stored in HDFS, \textit{mContain} uses Cerebral Cortex in its entirety. Cerebral Cortex is a open source big data cloud analytic platform that supports population-scale data analysis, visualization, model development, and intervention design for mobile sensor data. The development interface of Cerebral-Cortex provides all the capabilities of Apache Spark for efficient distributed computing and allows developers the opportunity to include their own custom routines. For development and debugging purposes, streams stored in HDFS can also be interfaced by the developers using Jupyter Notebook Environment. Algorithms responsible for contact tracing and crowd estimation are automated in jobs timed accordingly to compute and write derivative low frequency streams back into the HDFS through Cerebral-Cortex.  The phone can request these derivative streams back from the REST server using their respective UUIDs ($U_1$) and display the information onto the phone screen. 

In the following sections we focus our attention to the algorithmic processes, design and parameter choices responsible for efficient, distributed and accurate computation of contact tracing and social crowding.

\section{Cloud Algorithms - Authors Main Contributions to the \textit{mContain} project}
\label{sec:algorithms}
In this section we explain the algorithms in operation at different stages of data processing sequence responsible for computing the user and location specific derivative streams. \textbf{This section also encompasses the author's contributions in the the development of \textit{mContain} application in totality.}

\subsection{Proximity Encounter} We define an encounter between  a pair of users using Bluetooth beacons exchanged in between their smartphones. Although GPS traces provide the ability to estimate user location in terms of spatial coordinates, using it to pinpoint user with acceptable accuracy requires military grade sensor hardware not present in the smartphones. Thus Bluetooth serves as an better alternative for detecting proximity encounters. Ultrasound frequencies also present a possible sensor medium for detecting proximity encounters~\cite{mavroudis2017privacy}. However, due to the lower energy demands of BLE and its near ubiquitous availability in major smartphone categories, we decide to use Bluetooth signal as the principal means for estimating proximity exposure. Furthermore, both iOS and Android offers a distance estimate associated with inter device Bluetooth transmission. Let $u_i \in U$ be anonymized identity of the user present in the database. Each Bluetooth beacon exchange is defined by a distance function $b_{i,j}^{t} = d$. The beacon exchange function is defined by four parameters - $u_i$ and $u_j$ are the two users separated by a distance $d$ at time $t$. 

Let $T = \{t_1, t_{2}, t_{3}, ...., t_{n}\}$ represents the times of $n$ beacon exchanges, $b_{i,j}^{t_{m}}, 1<=m<=n$ between two users $u_i$ and $u_j$ in between two boundary time-points $t_1$ and $t_{n}$. Let $I(p,q)$ is an indicator function which is equal to $1$ if $p \le q$ and zero otherwise. Then, we define an encounter between the two users if two conditions are met. 

\begin{enumerate}
    \item First condition relates to the distance between the users and is defined as $\sum_{m=1}^{n} I(b_{i,j}^{t_{k+m}},\delta) \ge \mu $.
$\delta$ refers to the distance between the two users conducive for transmission of the virus and $\mu$ is the frequency of exchange. Thus, both users have to be within a $\delta$ distance of each other and should have at least $mu$ beacon exchanges between them.

    \item Second condition depends on the minimum time of contact between two users within the distance threshold $\delta$. Let $t_1 = \argmin_{t \in T} I(b_{i,j}^t,\delta)$ and $t_2 = \argmax_{t \in T} I(b_{i,j}^t,\delta)$. Then we consider the condition satisfied if $t_2 - t_1 \ge \tau$. $\tau$ refers to minimum time the two users need to be in proximity with each other for viral transmission. The two users have to be within distance $\delta$ for a minimum duration $\tau$.
\end{enumerate}
    
Thus our definition of proximity encounters between two users depend on three parameters $\delta$ and $\tau$ and $\mu$. The choice of $\delta$ and $\tau$ arise from Center for Diseases Control recommendations and taking into account the accuracy of distance estimation using Bluetooth. Although CDC recommends a minimum 6 feet distance between two socially distant users, $\delta$ is selected to be 12 metres which is the upper 95\% confidence bound based on our calculations from the findings of~\cite{5088918}. $\tau$ is selected to be 10 minutes based on the CDC recommendations since timestamps accurately identify the duration. Each pairwise encounter $e_i(u_1,u_2)$ between two users $u_1$ and $u_2$ is also represented by a GPS coordinate $s_i$ which is the mean latitude and longitude of all the GPS points sampled at each Bluetooth beacon exchange. We compute pairwise encounters in the cloud every one hour using all the beacon exchanges present within the hour.

\subsection{Social Crowding}

GPS trace produced by each encounter $e_i$ can be used to define a frequency distribution on top of a spatial coordinate system. We define a social crowding event by estimating the density of the distribution at specific places and designating the ones above a pre-selected threshold. Thus the problem of estimating crowding hot-spots translate to detecting the regions on the map with high densities. We use the DBSCAN clustering method which is the most commonly used density based clustering method employed in GPS processing literature~\cite{chatterjee2020smokingopp}. In DBSCAN clustering, clusters have the intrinsic property that each encounter in a specific cluster is within $\epsilon$ meters of at least one encounter belonging to the same cluster ($\epsilon$ = 50 metres). Clustering allows us to filter the sparse encounter locations. In each cluster the number of unique encounters per 10 square meter area determines the level of crowding in that location.

\subsubsection{DBSCAN Clustering:} 
DBSCAN requires two parameters - $\epsilon$ and the minimum number of points required to form a dense region ($n$). It starts with an arbitrary starting point that has not been visited. This point's $\epsilon$-neighborhood is retrieved, and if it contains sufficiently many points, a cluster is started. Otherwise, the point is labeled as noise. Note that this point might later be found in a sufficiently sized $\epsilon$-environment of a different point and hence be made part of a cluster. If a point is found to be a dense part of a cluster, its $\epsilon$-neighborhood is also part of that cluster. Hence, all points that are found within the $\epsilon$-neighborhood are added, as is their own $\epsilon$-neighborhood when they are also dense. This process continues until the density-connected cluster is completely found. Then, a new unvisited point is retrieved and processed, leading to the discovery of a further cluster or noise. We use the Manhattan distance as the distance estimate between two points since it is invariant to the earth rotation and provides a robust metric. DBSCAN clustering requires computation of a pairwise distance matrix which is memory expensive and requires additional approaches described in Section~\ref{sec:distributed} for efficient implementation.

\subsection{Contact Tracing}
\begin{figure*}[t]
    \centering
    \begin{subfigure}{0.33\textwidth}
        \centering
        \includegraphics[height=1.2in]{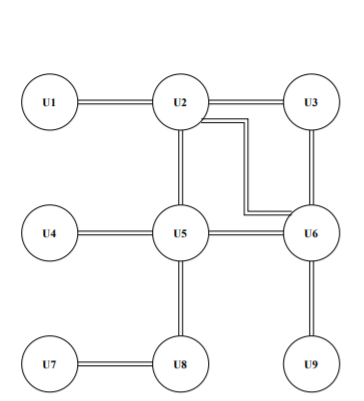}
        \caption{Initial Encounter}
        \label{fig:initial_encounter}
    \end{subfigure}%
    ~ 
    \begin{subfigure}{0.33\textwidth}
        \centering
        \includegraphics[height=1.2in]{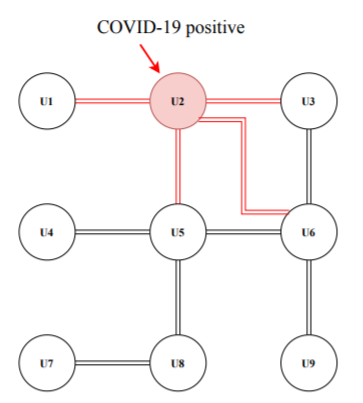}
        \caption{Postive Assessment}
        \label{fig:positive_assessment}
    \end{subfigure}
    ~
    \begin{subfigure}{0.3\textwidth}
        \centering
        \includegraphics[height=1.2in]{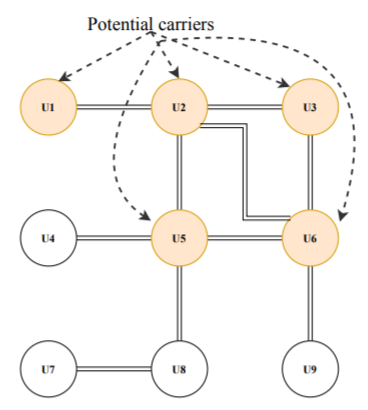}
        \caption{Exposed Participants}
        \label{fig:exposed}
    \end{subfigure}
    \caption{Automated Contact Tracing}
    \label{fig:contact_tracing}
\end{figure*}

Contact tracing is imperative for tracing exposure to people deemed COVID-19 positive. If any user and their testing agency agrees to share his/her positive COVID-19 result, we can query the pairwise proximity encounter table for determining other people who may have been positively exposed to the virus. Since each encounter has a timestamp associated with it, we can query within a past window based on when the symptoms start. Since the infection period of COVID-19 is determined to be 2 weeks, we select all the encounters going back at most 2 weeks from the symptom onset time. Figure~\ref{fig:contact_tracing} shows a first order contact tracing procedure using visual cues. Figure~\ref{fig:initial_encounter} shows the initial encounter graph with no COVID-19 positive user, each node represents a user, and each edge represents encounters within the last two weeks. Suppose user $U2$ tested positive for COVID-19 as shown in Figure~\ref{fig:positive_assessment}. Figure~\ref{fig:exposed} shows how application updates the status of those users who had an encounter with $U2$ as potential carrier.  The tracing process is automated and data-streams with positive exposure variable has associated access controls to shield the leakage of private information from any adversary in the server side. The system is only configured to generate first order tracing of affected participant to other people. However the routine is  to do traversal in the encounter graph based on the necessity.

\subsection{Information Discourse to User and Public}

The proximity encounters, crowding hot-spots are computed every hour of the day. The crowding hot-spots of the previous day are shown in the website using open street maps. However, users smartphone receives a real time overview of the crowding hot-spots every hour using custom rendering of the map view in the phone app.

In addition to the crowding hot-spots, users are notified once daily about their encounter counts and exposure to diagnosed COVID-19 patients. We notify users the amount of distinct encounters from the past day. Since proximity encounters are computed every hour, we avoid any duplicate encounters between two same users in the same location throughout the day and consider them only once. Users are also notified about the changes in proximity encounters from day to day. Specifically, any new encounters that are out of routine such as a new encounter partner is easily identified by the server and is communicated accordingly.  If the user has been in contact with any participant diagnosed with COVID-19, he/she will be sent notifications based on the final encounter table. 

Aggregate measures of social distancing can also be reliably estimated by computing metrics which communicate crowding events to the public. We identify two such metrics - average number of encounters per app user and number of encounters at the largest crowding event. These metrics are computed once daily and are shown in the \textit{mContain} website for public discourse of the information.

\subsection{Algorithmic Parallelization Using Apache Spark}
\label{sec:distributed}
The algorithmic routines running in an automated fashion needs to be efficient enough for scalability purposes. We make use of a 340 core cluster running Cerebral Cortex to effectively automate all the processes. The crux of the algorithmic efficiency depends on two methods alone. First the proximity encounters are calculated pairwise in a distributed manner. Thus, the quadratic complexity of computing encounters reduces by a significant margin depending on the resources available. The second bottleneck arises in the GPS clustering routine to determine social crowding events. Although GPS itself is sampled sparsely, in a sizeable metropolitan area the number of GPS points can grow exponentially large with addition of new users. Since DBSCAN clustering computes the pairwise distance matrix in the given set of points, this can easily become infeasible. Our approach towards avoiding this scenario lie in first computing the encounters first. Using only the mean estimate of the location for a single encounter can reduce the number of GPS points to a significant amount. 

Furthermore, in the event of a large user adoption, the system also has a parallel DBSCAN clustering routine. The parallel DBSCAN algorithm divides the whole GPS space into multiple sub-regions and performs clustering in each separate sub-region in a parallel fashion. In the final step we deal with merging the clusters of two nearby sub-regions depending on the radius of each cluster and the distance between two cluster centroids. This method reduces the memory footprint of a $O(n^2)$ distance matrix when $n$ is too large.

\section{Goal Resolution}
In this section, we discuss how \textit{mContain} app addresses the three design goals we have delineated above - Authenticity, Anonymity and Proper Communication with End User.

\emph{Authenticity:} When a person goes for testing for COVID-19, he fills up a form where he/she can enter the unique code from the app and gives consent for sharing the unique code with us if the test comes positive. If a diagnosed person is not the user of our app, he is asked to install the app so that we can get access to the unique code. As we obtain the test results directly from the testing agencies, it maintains the authenticity of the diagnosis of COVID-19 positive people in our system.  

\emph{Anonymity:} When a person installs the app, it assigns a unique random code to the app. We don't collect any user information. We are confident that there is no way anyone can link the code with a particular user. Also in case of proximity alert to a COVID-19 positive individual, we notify users without specifying any time of the day or location. Hence this also protects the identity of the infected individual.

\emph{Proper Communication with End User:} As discussed we have two different alerts, one being crowding alert and another one is contact alert. Our algorithm calculates the crowding hot-spots using GPS and Bluetooth data. We alert the users if they have been in any of those crowding hot-spots for more than 10 minutes. As for contact alerts, we use Bluetooth to detect the proximity of any app user to a COVID-19 diagnosed individual. If any user has been within 6 feet with the infected individual for more than cumulative 10 minutes throughout the day, we alert the users saying that they have been in contact with a COVID-19 positive individual.

\section{Limitations}
Although \textit{mContain} makes use of anonymous codes for app users, the sensitive information being used by the application including the location and proximity of users puts a limit on the trustworthiness of the application to the general public. \textit{mContain} makes choices in the utility privacy trade-off curve to make the most use of concurrent technological instruments for better use. The significant limitations of the application mostly lie in the social crowding arena. The crowding hot-spots being shown in the application need to be personalized for ensuring the sanctity of private locations such as other users home. The website also needs to only show public places and respect the users private location. Although technically possible, this required significant time to annotate and innovate on the application goals. The limited adoption of the application can also be attributed to the unavailability of the application in Apple store. However, the necessity of applications like \textit{mContain} can be easily deduced from the news of a collaborative effort between Apple and Google to make their operating systems capable of mentioned technologies. In any case the limitations are a source of education going forward. They serve as invaluable lessons for future developers, innovators and technological applications. 

\section{Discussion}
\textit{mContain} was developed as an emergency measure in times of extraordinary crisis facing the world around us. The goal was to raise awareness and help the residents of Memphis metropolitan area and greater Shelby County through use of common mobile technologies available to them.
Although the initial target of releasing the application in both iOS and android fell short due to the privacy constraints of Apple app store, we have gained invaluable experience in envisaging, deploying and sparse adoption of the application in Google Play Store. Irrespective of the intentions, an application that uses sensitive data from human subjects have to be more aligned with the privacy constraints. Nevertheless, the privacy limitations of \textit{mContain} can be significantly reduced with further development and the lessons learnt can be a stepping stone for a future use case. 

\textit{mContain} sheds light on the promise of mobile sensor technology to help people in uncertain times of crisis. Proximity encounters can successfully lead to people's awareness of possible viral exposure. Contact tracing technology has been successfully used by multiple countries to help understand the possible transmission pathway of the virus.  

From a technical and computational point of view, \textit{mContain} makes successful use of distributed big data computing for scalable and efficient application of mobile sensor technologies. The backend of \textit{mContain} built using Cerebral Cortex and Apache Spark could accommodate emergency development cycle of only a few weeks and host a software aimed at a population scale deployment.

Overall, We acknowledge everyone who worked tremendously hard and contributed to making \textit{mContain} a reality. The significant new understanding of the health application world, privacy constraints and regulatory limitations definitely make us better prepared in the event of a future viral pandemic. 
Fortunately, with incredible efforts from scientists and essential workers all over the world, at present COVID-19 vaccines are available to people and we are moving towards better future one step at a time.

\bibliographystyle{splncs04}
\bibliography{samplepaper}

\begin{thebibliography}{10}
\providecommand{\url}[1]{\texttt{#1}}
\providecommand{\urlprefix}{URL }
\providecommand{\doi}[1]{https://doi.org/#1}

\bibitem{gov.il}
Hamagen - the ministry of health app for fighting the spread of coronavirus,
  \url{https://govextra.gov.il/ministry-of-health/hamagen-app/download-en/}

\bibitem{rss}
Unacast updates social distancing scoreboard - unacast,
  \url{https://www.unacast.com/post/unacast-updates-social-distancing-scoreboard}

\bibitem{apple_newsroom_2021}
Apple releases new covid-19 app and website based on cdc guidance (Mar 2021),
  \url{https://www.apple.com/newsroom/2020/03/apple-releases-new-covid-19-app-and-website-based-on-CDC-guidance/}

\bibitem{sharonbegleyapril_2020}
April, S.B.: New digital tools could speed up covid-19 contact tracing (Apr
  2020),
  \url{https://www.statnews.com/2020/04/02/coronavirus-spreads-too-fast-for-contact-tracing-digital-tools-could-help/}

\bibitem{armbruster2007contact}
Armbruster, B., Brandeau, M.L.: Contact tracing to control infectious disease:
  when enough is enough. Health care management science  \textbf{10}(4),
  341--355 (2007)

\bibitem{browne2015modeling}
Browne, C., Gulbudak, H., Webb, G.: Modeling contact tracing in outbreaks with
  application to ebola. Journal of theoretical biology  \textbf{384},  33--49
  (2015)

\bibitem{chatterjee2020smokingopp}
Chatterjee, S., Moreno, A., Lizotte, S.L., Akther, S., Ertin, E., Fagundes,
  C.P., Lam, C., Rehg, J.M., Wan, N., Wetter, D.W., et~al.: Smokingopp:
  Detecting the smoking'opportunity'context using mobile sensors. Proceedings
  of the ACM on Interactive, Mobile, Wearable and Ubiquitous Technologies
  \textbf{4}(1),  1--26 (2020)

\bibitem{danquah2019use}
Danquah, L.O., Hasham, N., MacFarlane, M., Conteh, F.E., Momoh, F., Tedesco,
  A.A., Jambai, A., Ross, D.A., Weiss, H.A.: Use of a mobile application for
  ebola contact tracing and monitoring in northern sierra leone: a
  proof-of-concept study. BMC infectious diseases  \textbf{19}(1),  1--12
  (2019)

\bibitem{dixon2015contact}
Dixon, M.G., Taylor, M.M., Dee, J., Hakim, A., Cantey, P., Lim, T., Bah, H.,
  Camara, S.M., Ndongmo, C.B., Togba, M., et~al.: Contact tracing activities
  during the ebola virus disease epidemic in kindia and faranah, guinea, 2014.
  Emerging infectious diseases  \textbf{21}(11), ~2022 (2015)

\bibitem{farrahi2014epidemic}
Farrahi, K., Emonet, R., Cebrian, M.: Epidemic contact tracing via
  communication traces. PloS one  \textbf{9}(5),  e95133 (2014)

\bibitem{ferguson2020impact}
Ferguson, N.M., Laydon, D., Nedjati-Gilani, G., Imai, N., Ainslie, K.,
  Baguelin, M., Bhatia, S., Boonyasiri, A., Cucunub{\'a}, Z., Cuomo-Dannenburg,
  G., et~al.: Impact of non-pharmaceutical interventions (npis) to reduce
  covid-19 mortality and healthcare demand. imperial college covid-19 response
  team. Imperial College COVID-19 Response Team p.~20 (2020)

\bibitem{ha2016evaluation}
Ha, Y.P., Tesfalul, M.A., Littman-Quinn, R., Antwi, C., Green, R.S., Mapila,
  T.O., Bellamy, S.L., Ncube, R.T., Mugisha, K., Ho-Foster, A.R., et~al.:
  Evaluation of a mobile health approach to tuberculosis contact tracing in
  botswana. Journal of health communication  \textbf{21}(10),  1115--1121
  (2016)

\bibitem{hamilton_2020}
Hamilton, I.A.: Compulsory selfies and contact-tracing: Authorities everywhere
  are using smartphones to track the coronavirus, and it's part of a massive
  increase in global surveillance (Apr 2020),
  \url{https://www.businessinsider.com/countries-tracking-citizens-phones-coronavirus-2020-3#south-korea-gives-out-detailed-information-about-patients-whereabouts-1}

\bibitem{hernandez2020evaluating}
Hern{\'a}ndez-Orallo, E., Manzoni, P., Calafate, C.T., Cano, J.C.: Evaluating
  how smartphone contact tracing technology can reduce the spread of infectious
  diseases: the case of covid-19. IEEE Access  \textbf{8},  99083--99097 (2020)

\bibitem{holmes_2020}
Holmes, A.: Singapore is using a high-tech surveillance app to track the
  coronavirus, keeping schools and businesses open. here's how it works. (Mar
  2020),
  \url{https://www.businessinsider.com/singapore-coronavirus-app-tracking-testing-no-shutdown-how-it-works-2020-3}

\bibitem{ilyushina_2020}
Ilyushina, M.: How russia is using authoritarian tech to curb coronavirus (Mar
  2020),
  \url{https://www.cnn.com/2020/03/29/europe/russia-coronavirus-authoritarian-tech-intl/index.html}

\bibitem{kiss2005disease}
Kiss, I.Z., Green, D.M., Kao, R.R.: Disease contact tracing in random and
  clustered networks. Proceedings of the Royal Society B: Biological Sciences
  \textbf{272}(1570),  1407--1414 (2005)

\bibitem{kretzschmar2020impact}
Kretzschmar, M.E., Rozhnova, G., Bootsma, M.C., van Boven, M., van~de Wijgert,
  J.H., Bonten, M.J.: Impact of delays on effectiveness of contact tracing
  strategies for covid-19: a modelling study. The Lancet Public Health
  \textbf{5}(8),  e452--e459 (2020)

\bibitem{lipsitch2003transmission}
Lipsitch, M., Cohen, T., Cooper, B., Robins, J.M., Ma, S., James, L.,
  Gopalakrishna, G., Chew, S.K., Tan, C.C., Samore, M.H., et~al.: Transmission
  dynamics and control of severe acute respiratory syndrome. science
  \textbf{300}(5627),  1966--1970 (2003)

\bibitem{mackensen2012bluetooth}
Mackensen, E., Lai, M., Wendt, T.M.: Bluetooth low energy (ble) based wireless
  sensors. In: SENSORS, 2012 IEEE. pp.~1--4. IEEE (2012)

\bibitem{5088918}
{Mascetti}, S., {Bettini}, C., {Freni}, D., {Wang}, X.S., {Jajodia}, S.:
  Privacy-aware proximity based services. In: 2009 Tenth International
  Conference on Mobile Data Management: Systems, Services and Middleware. pp.
  31--40 (2009). \doi{10.1109/MDM.2009.14}

\bibitem{mavroudis2017privacy}
Mavroudis, V., Hao, S., Fratantonio, Y., Maggi, F., Kruegel, C., Vigna, G.: On
  the privacy and security of the ultrasound ecosystem. Proceedings on Privacy
  Enhancing Technologies  \textbf{2017}(2),  95--112 (2017)

\bibitem{mitter_2020}
Mitter, S.: Government launches multi-language coronavirus tracking app,
  aarogya setu (Apr 2020),
  \url{https://yourstory.com/2020/04/india-government-coronavirus-tracking-app-aarogya-setu}

\bibitem{porco2004logistics}
Porco, T.C., Holbrook, K.A., Fernyak, S.E., Portnoy, D.L., Reiter, R.,
  Arag{\'o}n, T.J.: Logistics of community smallpox control through contact
  tracing and ring vaccination: a stochastic network model. BMC public health
  \textbf{4}(1),  1--20 (2004)

\bibitem{sacks2015introduction}
Sacks, J.A., Zehe, E., Redick, C., Bah, A., Cowger, K., Camara, M., Diallo, A.,
  Gigo, A.N.I., Dhillon, R.S., Liu, A.: Introduction of mobile health tools to
  support ebola surveillance and contact tracing in guinea. Global Health:
  Science and Practice  \textbf{3}(4),  646--659 (2015)

\bibitem{swanson2018contact}
Swanson, K.C., Altare, C., Wesseh, C.S., Nyenswah, T., Ahmed, T., Eyal, N.,
  Hamblion, E.L., Lessler, J., Peters, D.H., Altmann, M.: Contact tracing
  performance during the ebola epidemic in liberia, 2014-2015. PLoS neglected
  tropical diseases  \textbf{12}(9),  e0006762 (2018)

\bibitem{thunstrom2020benefits}
Thunstr{\"o}m, L., Newbold, S.C., Finnoff, D., Ashworth, M., Shogren, J.F.: The
  benefits and costs of using social distancing to flatten the curve for
  covid-19. Journal of Benefit-Cost Analysis  \textbf{11}(2),  179--195 (2020)

\bibitem{wu2006reducing}
Wu, J.T., Riley, S., Fraser, C., Leung, G.M.: Reducing the impact of the next
  influenza pandemic using household-based public health interventions. PLoS
  Med  \textbf{3}(9), ~e361 (2006)

\bibitem{yasaka2020peer}
Yasaka, T.M., Lehrich, B.M., Sahyouni, R.: Peer-to-peer contact tracing:
  development of a privacy-preserving smartphone app. JMIR mHealth and uHealth
  \textbf{8}(4),  e18936 (2020)

\end{thebibliography}

\end{document}